\documentclass{aa}
\usepackage{graphicx}
\usepackage{times}
\usepackage{psfig}
\usepackage{natbib}
\bibpunct{(}{)}{;}{a}{}{,}

\def\hide#1{}

\def\mpcoh{{\,h^{-1}\,\rm Mpc}}
\def\logten{\log_{10}}

\begin{document}

\title{Galaxy clustering from COMBO-17: The halo occupation
distribution at $\langle z\rangle=0.6$} 
\author{S. Phleps\inst{1,4}, J.~A. Peacock\inst{1} \and
K. Meisenheimer\inst{2} \and C. Wolf\inst{3} }
\offprints{S.~Phleps (sphleps@mpe.mpg.de)}

\institute{
Institute for Astronomy, University of Edinburgh, Royal Observatory, Blackford Hill, Edinburgh EH9 3HJ, UK
\and Max-Planck-Institut f\"ur Astronomie, K\"onigstuhl 17, D-69117
Heidelberg, Germany
\and Department of Physics, University of Oxford, Denys Wilkinson Building., Keble Road, Oxford OX1 3RH, UK
\and Max-Planck-Institut f\"ur Extraterrestrische Physik, Giessenbachstra\ss e, D-85748 Garching, Germany
}

\titlerunning{Galaxy clustering from COMBO-17}
\date{}
\date{Received 14. 06. 2005 / Accepted 26. 06. 2006}

\abstract{We present measurements of galaxy clustering at redshift
$\langle z \rangle=0.6$ using $10\,360$ galaxies with photometric redshifts 
over an area of 0.78 deg$^2$ from the COMBO-17 survey. To obtain a result
that is unaffected by redshift uncertainties, we calculate the
projected correlation function $w(r_p)$, giving results for red sequence and blue
cloud galaxies separately. The correlation function of the red galaxies displays
clear deviations from a power law at comoving separations
around 1 to $3\mpcoh$, and similar but weaker trends are suggested
by the  data for the blue galaxies. To interpret these results, we fit the correlation
functions with analytical predictions derived from a simple halo occupation model.
This combines linear clustering of the underlying mass with 
a description of the number of galaxies occupying each dark-matter halo
(the halo occupation distribution). 
If the occupation numbers are taken to be a simple
power law $N \propto \smash{M^\alpha}$, then $\alpha \simeq 0.5$ and
$\alpha \simeq 0.2$ for red and blue galaxies respectively.
These figures are little different from the values required to fit
present-day clustering data. The power-spectrum shape is assumed
to be known in this exercise, but we allow the data to determine
the preferred value of $\sigma_8$, the linear power-spectrum normalization.
The average normalization inferred from
red and blue galaxies at $\langle z \rangle=0.6$
is $\sigma_8=1.02\pm0.17$ at zero redshift,
consistent with independent estimates of this local value.
This agreement can be regarded as a verification of the
hierarchical growth of the halo mass function.

}
\maketitle
\section{Introduction}

In current models of galaxy formation, structure grows hierarchically
from small Gaussian density fluctuations. Galaxies are presumed to form
within virialized dark matter haloes when the baryonic gas
cools and condenses into stars (e.g. \citealp{Cole00}).
The formation and evolution of galaxies should thus be
closely tied to the merging history of dark matter haloes.
This paper uses measurements of galaxy clustering at intermediate
redshift to test this basic picture.

The complex relation between galaxies and dark matter 
has become clearer only slowly. Empirically,
galaxies display biased clustering in which the
amplitude of their correlations varies with galaxy type:
older galaxies are generally much more strongly
clustered than young, starforming galaxies, and bright galaxies are more
strongly clustered than faint galaxies
(e.g. \citealp{Davis76,Norberg02,PhlepsMeise03}).
It is generally believed that such trends can be understood through
the tendency for dark-matter haloes to display clustering that is
larger for rare massive haloes (e.g. 
\citealp{ColeKaiser89,Mo96,ShethTormen99}).

However, a long-standing challenge has been to understand how these ideas could be
implemented in the context of Cold Dark Matter (CDM) models. The
galaxy correlation function has long been known  to be extremely close to
a single power law (\citealp{Totsuji69}, \citealp{PeeblesPowerLaw}), and yet
this is not the case for the nonlinear mass correlations in a CDM
model. 
Here, the matter correlation function rises above a best-fit power law on
scales $r\la 1 \mpcoh$
and falls below it again on scales $r\la 0.2 \mpcoh$
(\citealp{Jenkins98} and references therein). This puzzle was only resolved
when it became clear that the correlation function of dark matter haloes
(including subhaloes inside large host haloes) differs significantly from the correlation
function of the mass. In practice, the predicted correlation function of galaxy-scale haloes
follows a power law down to $100\,h^{-1}$\,kpc, with an amplitude and slope similar to the
data on real galaxies \citep{Kravtsov99,Neyrinck04,Kravtsov04,Tasitsiomi04}. This
phenomenon underlies the considerable scale-dependent bias
predicted by semianalytic and hydrodynamic simulation models of galaxy formation
(\citealp{Colin99,Kauffmann99I,Pearce99,Benson00,CenOstriker00,Somerville01,Yoshikawa01,Weinberg04}).

These developments in turn stimulated a simpler and more direct insight into 
bias and its dependence on scale, through the so-called halo model
(e.g. \citealp{Jing98,Seljak00,Peacock2000,CooraySheth02} and references therein).
Here, the shape of the correlation function is determined by the
linear clustering of the dark matter, and the relation of the galaxies to
the dark matter halos in which they reside (the Halo Occupation Distribution;
HOD).
In particular, a break in slope is expected when the correlation
function changes from being dominated by pair counts of galaxies in
separate dark matter halos to the small-scale regime, where pairs come
from two galaxies that reside in the same halo. Any pure power law
correlation function would require coincidental alignment of these two terms,
and indeed analyses of the two-point correlation function of galaxies
in the local universe have detected small deviations from the
power-law form (\citealp{Hawk03,Zehavi04,Zehavi05,Abazajian05}).

In this paper, we use the COMBO-17 survey \citep{COMBOMain04} to carry
out a similar investigation of the exact shape of the correlation function at
higher redshifts.
We calculate the projected correlation function $w(r_p)$ for red sequence
and blue cloud COMBO-17 galaxies (following the definition of
\citealp{Bell03}), in the redshift bin $0.4<z<0.8$. By comparing these results
to the predictions of the halo model, we are able to infer the mean number of galaxies
per halo of a given mass (the halo occupation number) and also
the $z=0$ power-spectrum normalization $\sigma_8$ (the rms density variation
averaged over $8 h^{-1}$ Mpc spheres).

This paper is structured as follows: The COMBO-17 survey and the data
used in this analysis are briefly
described in Sect. \ref{COMBO}. 
The halo model is introduced in Sect. \ref{HaloModel}. 
The method used to estimate the
projected correlation function is explained in Sect. \ref{Method}. 
In Sect. \ref{Results} we investigate the shape of the correlation
function for red sequence and blue cloud galaxies, and in Sect. \ref{Discussion}  the
results are discussed.
We assume a cosmological geometry taken from the WMAP results 
(\citealp{Spergel03,Spergel06})
and the final 2dFGRS power spectrum results (\citealp{Cole05}):
a flat model with $\Omega_m=0.25$. All lengths quoted are in comoving units.
Normally, we show explicit dependence on $h$ (which denotes $H_0 / 100 \rm \,km\,s^{-1}\,Mpc^{-1}$);
but for absolute magnitudes we suppress this dependence, so that
$M_B$ denotes $M_B - 5 \log_{10}h$.

\section{Data base: The COMBO-17 Survey}\label{COMBO}
To date, COMBO-17 ({\bf C}lassifying {\bf O}bjects with {\bf M}edium
{\bf B}and {\bf O}bservations in 17 filters) has surveyed three
disjoint $\sim 31'\times 30'$ 
southern equatorial fields (for their coordinates see \citealp{Wolf03})
to deep limits in $5$ broad and $12$ 
medium passbands, covering wavelengths from $400$ to $930$\,nm. A
detailed description of the survey along with filter
curves can be found in \citet{COMBOMain04}.
All observations were carried out using the Wide Field Imager at the
MPG/ESO 2.2 m-telescope on La Silla, Chile.

In each filter, typically $10$ to $20$ individual exposures were taken
(up to $50$ for ultradeep $R$-band images totalling $20$\,ks with seeing $\la 0\farcs8$).
Galaxies were detected on the deep $R$-band images by using {\bf SE}xtractor
\citep{Bertin96}. The spectral energy distributions (SEDs) for $R$-band detected objects
were measured by performing seeing-adaptive, weighted-aperture
photometry in all $17$ frames at the position of the $R$-band detected
object. All magnitudes are quoted with a Vega zero point.

Using the 17-band photometry, objects are classified  using a scheme
based on template spectral energy distributions
\citep{Wolf01a,Wolf01b}. The classification algorithm basically
compares the observed colours of each object with a colour library of
known objects. This colour library is assembled from observed and model spectra
by synthetic photometry performed using an accurate representation of the instrumental
characteristics of COMBO-17. For galaxy classification, we use
P\'{E}GASE model spectra (see \citealp{Pegase97} for an earlier version of the model).
The template spectra are a two-dimensional
age/reddening sequence, in which a fixed exponential star formation
timescale $\tau = 1$\,Gyr is assumed, ages vary between $50$\,Myr and $15$\,Gyr,
and the reddening $E(B-V)$ can be as large as $0.5$\,mag, adopting a Small
Magellanic Cloud Bar extinction curve. 
Note that we do not apply any morphological star/galaxy separation or
use other criteria.

Using a minimum variance estimator, each
object is assigned a redshift (if it is
not classified as a star).  The  redshift errors in this
process depend on magnitude and type of the object, and for 
galaxies can be approximated by
\begin{eqnarray}\label{redshifterrors}
\frac{\sigma_z}{(1+z)}=0.007 \left( 1+10^{0.8(R-21.6)} \right)^{1/2}~.
\end{eqnarray}
The galaxy redshift estimate quality has been tested by comparison
with spectroscopic redshifts for almost 1000 objects (see \citealp{COMBOMain04}).
At bright limits $R < 20$, the redshifts are accurate to
$\sigma_z/(1 + z) \simeq 0.01$, and the error is dominated by mismatches between
template and real galaxy spectra. This error can contain a systematic component
that is dictated by the exact filter placement, but these `redshift focusing'
effects are of the order of magnitude of the random redshift errors for $z<1$ and are 
unimportant for the current analysis.
At the median apparent magnitude
$R \simeq 23$, $\sigma_z/(1 + z) \sim 0.02$. For the faintest
galaxies, the redshift accuracy 
approaches those achievable using traditional broadband photometric
surveys, $\sigma_z/(1 + z)  \ga 0.05$. We thus restricted our analysis
to galaxies with $I< 23$.

\begin{figure}[h]
\centerline{\psfig{figure=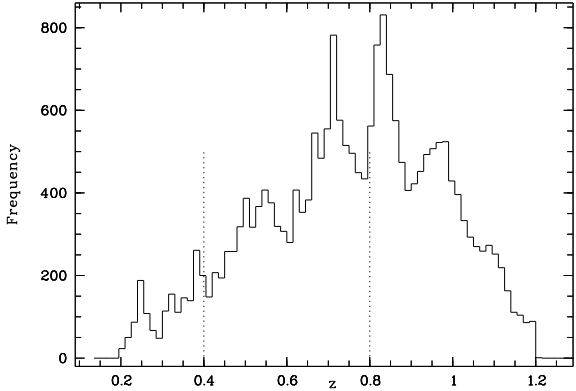,angle=270,clip=t,width=8.3cm}}
\caption[ ]{The redshift distribution for galaxies with $I<23$ and $M_B < -18$  in the three
COMBO-17 fields. The dotted lines indicate the redshift range
over which we define a volume-limited sample for the clustering analysis.
\label{zdistrib}}
\end{figure}

Fig. \ref{zdistrib} shows the redshift distribution of the $22\,310$ COMBO-17
galaxies between $z = 0.2$ and $z = 1.2$ (with $I < 23$ and $M_B < -18$).
The peak at $z=0.733$ in Fig. \ref{zdistrib} is due to a real structure
in the Chandra Deep Field South, which has been spectroscopically confirmed \citep{Gilli03}.  
In order to define a volume limited sample, we restrict our analysis to
the redshift range $0.4 < z < 0.8$ and galaxies brighter than
$M_B= -18$, which leaves us with $10\,360$ galaxies for the analysis.
Note that $B$-band luminosities can be
determined directly without any
$K$-correction uncertainty, based on the photometry in our 17 filters between $400$ and
$930$\,nm and an interpolation of the corresponding template
spectra. We do not apply any evolutionary corrections.

\begin{figure}[h]
\centerline{\psfig{figure=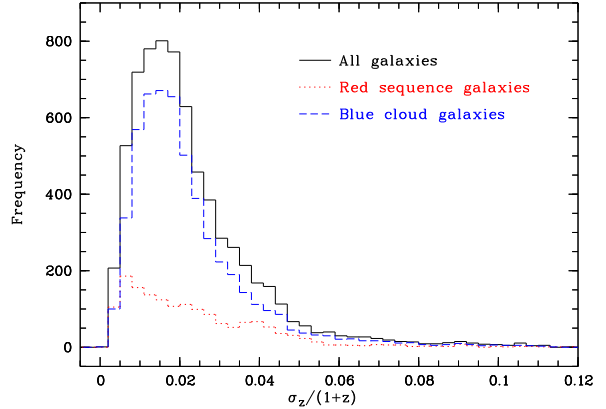,angle=270,clip=t,width=8.3cm}}
\caption[ ]{Distribution of relative redshift errors of galaxies with
$I<23$, $M_B < -18$ and $0.4 < z < 0.8$ (solid line). Also
shown are the redshift error distributions for red sequence and blue
cloud galaxies.
\label{relzerror}}
\end{figure}

The distribution of the redshift errors
for all galaxies in our subsample with $I<23$, $M_B < -18$ 
and $0.4 < z < 0.8$ is shown in Fig. \ref{relzerror}.
We use the prescription of \citet{Bell03} to separate galaxies into the
red-sequence component and the remaining blue cloud component:
\begin{equation}\label{redbluedef1}
{\rm Red\ sequence:\quad} (U-V) > (U-V)_{\mathrm {lim}}
\end{equation}
\begin{equation}\label{redbluedef2}
{\rm Blue\ cloud:\quad} (U-V) < (U-V)_{\mathrm {lim}} 
\end{equation}
\begin{equation}\label{redbluedef3}
(U-V)_{\mathrm {lim}} = 1.25-0.4 z -0.08(M_V-5\logten h+20)~,
\end{equation}
where $z$ denotes the redshift of each single galaxy. 

Note that the cut that separates the red sequence galaxies from the blue ones depends
on both redshift and absolute $V$ magnitude.
This yields 2404 and 7956 galaxies in the red and blue subsamples;
the former tend to have more accurate redshifts, as shown in
Fig. \ref{relzerror}. This is not because the classification scheme
works better for the red galaxies, but because they are on average
brighter than the blue ones.

Table \ref{numbertab} shows the number of red sequence
and blue cloud galaxies per COMBO-17 field.

\begin{table}
\centering
\caption[ ]{The number of red sequence
and blue cloud galaxies (by the definition of equations \ref{redbluedef1},
\ref{redbluedef2} and \ref{redbluedef3}), with $I<23$, $M_B < -18$  
and $0.4 < z < 0.8$   
per COMBO-17 field.\\\label{numbertab}}
\begin{tabular}{c|r r   }
COMBO-17 field& $N_{\rm red}$& $N_{\rm blue}$\\ \hline
CDFS&$752$&$2782$\\
A901&$836$&$2563$\\
S11&$816$&$2611$\\
\end{tabular}
\end{table}

\section{The halo model of galaxy clustering}\label{HaloModel}

In discussing the results of clustering analyses of the COMBO-17 data,
we will make frequent comparisons with theoretical predictions. We
therefore now summarise the framework used to carry out this modelling. 
This goes back to the paradigm introduced by \citet{WhiteRees78}: 
galaxies form through the cooling of baryonic material in
virialized haloes of dark matter. 
The mass function and density profiles of these haloes can
be expressed in terms of simple fitting formulae derived from
N-body simulations, and the large-scale clustering of haloes can
be derived analytically for Gaussian density fields.
This concentration on dark-matter haloes gives concrete form to
earlier work on the clustering statistics generated by
distributions of extended clumps (see e.g.
\citealp{NeymanScott52} and \citealp{ScherrerBertschinger91}).

With an accurate description of dark-matter clustering to hand,
the stage was set for an extension to galaxies via the `halo model'
(\citealp{MaFry00,Seljak00,Peacock2000,CooraySheth02}).
In this picture, the key remaining uncertainty is the way in
which galaxies occupy the dark matter haloes; this can be
regarded as an unknown function, to be probed experimentally.
The halo model then allows a simple and direct understanding of
many features of galaxy clustering, and how the clustering of galaxies
differs from that of the mass.

In this approach, the density field is a
superposition of dark-matter haloes, with small-scale clustering arising
from neighbours in the same halo. The corresponding real-space correlation function
can be written as a combination of two parts:
\begin{eqnarray}
\xi_r = \xi_{\rm lin} + \xi_{\rm halo}~,
\end{eqnarray}
the first term representing the clustering of the dark matter haloes,
and the second correlations from within a single halo. 
Large-scale halo correlations depend on mass, and the linear bias
parameter for a given class of
haloes, $b(M)$, depends on the rareness of
the fluctuation and the rms of the underlying field
\citep{Kaiser84,ColeKaiser89,Mo96,ShethTormen99}, usually measured in spheres of $8
\mpcoh$ and termed $\sigma_8$. The mass profile of the haloes is known
from simulations, and may be assumed to follow either an NFW profile
\citep{NFW1,NFW2,NFW3}, or a \citet{Moore99} profile.

The key feature that allows bias to be included is to encode
all the complications of  galaxy formation via
an halo occupation number: the number of galaxies
found above some luminosity threshold in a virialized halo of a given
mass $M$. A simple but instructive model for this  halo occupation distribution (HOD) is
\begin{eqnarray}\label{HODeqn}
N(M) = \cases{
0\quad &($M<M_c$)\cr
(M/M_c)^{\alpha}\quad &($M>M_c$)~.\cr
}
\end{eqnarray}
This is closely related to the mass-dependent weight introduced
by \citet{Jing98b}.
A model in which light traces mass exactly would have $M_c \rightarrow 0$ and $\alpha=1$.
The galaxies are assumed to be split into a central galaxy plus some
number of satellite galaxies, which follow the mass distribution in the halo.
It is necessary to make an assumption about the statistics of the
HOD -- in particular whether $N$ is a causal function of $M$ or whether
it obeys a Poisson distribution. It is known that sensible results
require sub-Poisson behaviour, and we assume the extreme limit in
which $N$ is perfectly determined by $M$.
Putting all these ingredients together, the galaxy correlation function can
be calculated analytically. 

Since its initial development, the halo model has been applied
successfully to the interpretation of the correlation function of
galaxies in the local universe,
notably by \citet{Zehavi04}, who detected an inflection in the correlation
function of SDSS red galaxies, interpreting it as indicating the
transition regime between clustering dominated by 1-halo and 2-halo terms.
The occupation model has been elaborated quite significantly
(e.g.
\citealp{Abazajian05,Zheng05,Kravtsov04,Tasitsiomi04,Zentner05}),
including up to three parameters 
for the occupation distribution, plus the inclusion of some nonlinear
evolution of the power spectrum in the $\xi_{\rm lin}$ term.  These
sophistications can improve the detailed fit to correlation-function
data from simulations, but they do run somewhat counter to the
original heuristic spirit of the model. In this work, we shall retain
the original method of calculation, as 
described in detail by \citet{Peacock2000},
together with the simple power-law occupation model. This approach
seems justifiable in a first exploration of intermediate-redshift clustering,
and the main features of interest are in any case relatively
robust. 

The form of any transition-regime feature in the correlation function
depends mainly on the mean halo mass occupied by the galaxies,
-- i.e. the average value of $M$, weighted by $N(M)$ --
and is insensitive to the details of their distribution within the
halo. This typical halo mass is determined by our two-parameter
model for $N(M)$ plus the halo mass function. We shall use the
additional constraint of the observed number density of galaxies
under study, so that there remains a single free parameter in 
the model -- which we take to be the occupation slope $\alpha$.
For a given value of $\alpha$, the number density determines the
cutoff mass and hence the average halo mass.
Different models of the HOD can of course be used; Zehavi et al.
(2004) take $N=1$ above $M_c$ to represent central galaxies,
plus a power-law $N(M)$ representing satellites, which commences
at a mass of approximately $20M_c$ with a slope of $\alpha\simeq 1$.
In practice, this model gives results similar to the single
power law with $\alpha \simeq 0.6$, showing that
the detailed shape of the HOD is hard to measure given only correlation-function data.
The typical number-weighted halo mass is a more robust quantity,
and this is probably the best way to compare different HOD models.

We will normally assume a standard
flat cosmology with $\Omega_m=0.25$, $\Omega_b=0.045$
and $h=0.73$ where $H_0=100 h $\,km s$^{-1}$ Mpc$^{-1}$
(\citealp{Spergel06,Cole05}).
The most uncertain cosmological parameter is the 
normalization of the power spectrum, $\sigma_8$.
We will often assume a standard value of $\sigma_8=0.9$, but it
is also of interest to 
leave both $\sigma_8$ and the power-law index $\alpha$ of the halo
occupation number (equation \ref{HODeqn}) as free parameters, and
determine them from a fit to the measured galaxy correlation functions.

\section{Redshift space correlations}\label{Method}

\subsection{Projected correlations}

With a sample of the present size, only the two-point
correlation function, $\xi(r)$, can be determined accurately.
Even this is not straightforward, because of the need to
work in redshift space. While it is possible to measure angular
positions on the sky with high precision, peculiar velocities as well
as redshift errors distort the 
galaxy pattern along the line of sight, making $\xi({\bf r})$
appear anisotropic, and tending to reduce its amplitude.

These problems can be overcome by splitting the separation vector ${\bf r}$
of a pair of objects into components lying on the plane of sky,
$r_p$, and along the line of sight, $\pi$, and compute the correlation
function $\xi(r_p,\pi)$ as a function of these two
components. Projecting $\xi(r_p,\pi)$  onto the $r_p$ axis gives the
projected function $w(r_p)$, which is independent of any radial
distortions \citep{DavisPeebles83}. For small angles $r^2=r_p^2+\pi^2$.
Thus the projected correlation function is defined as
\begin{eqnarray}\label{projection}
w(r_p)&=& \int_{-\infty}^\infty{\xi\left[(r_p^2+\pi^2)^{1/2}\right]~{\mathrm
d}\pi}\nonumber\\
&=&2\int_{r_p}^\infty{\xi(r)(r^2-r_p^2)^{-1/2} \, r~{\mathrm d}r}~.
\end{eqnarray}
Note that $w(r_p)$ has dimensions of length.
Since the correlation function converges rapidly to
zero with increasing pair separation, the integration limits
do not have to be $\pm \infty$, but they
have to be large enough to include all correlated pairs. As will
be explained in the following paragraph, this is a crucial point when the
redshift errors are large.

\begin{figure}[ht]
\centerline{\psfig{figure=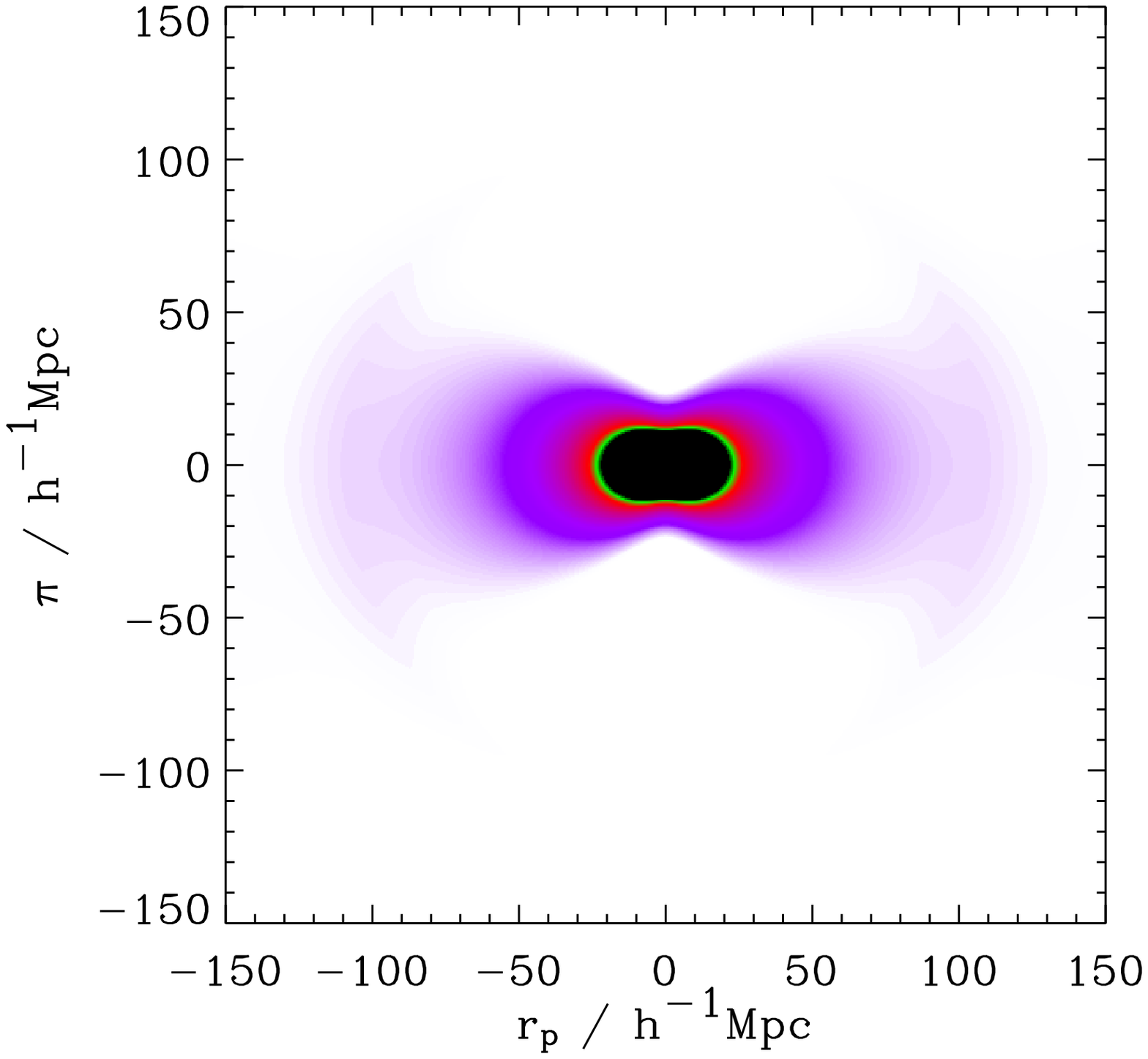,angle=0,clip=t,width=8.cm}}
\centerline{\psfig{figure=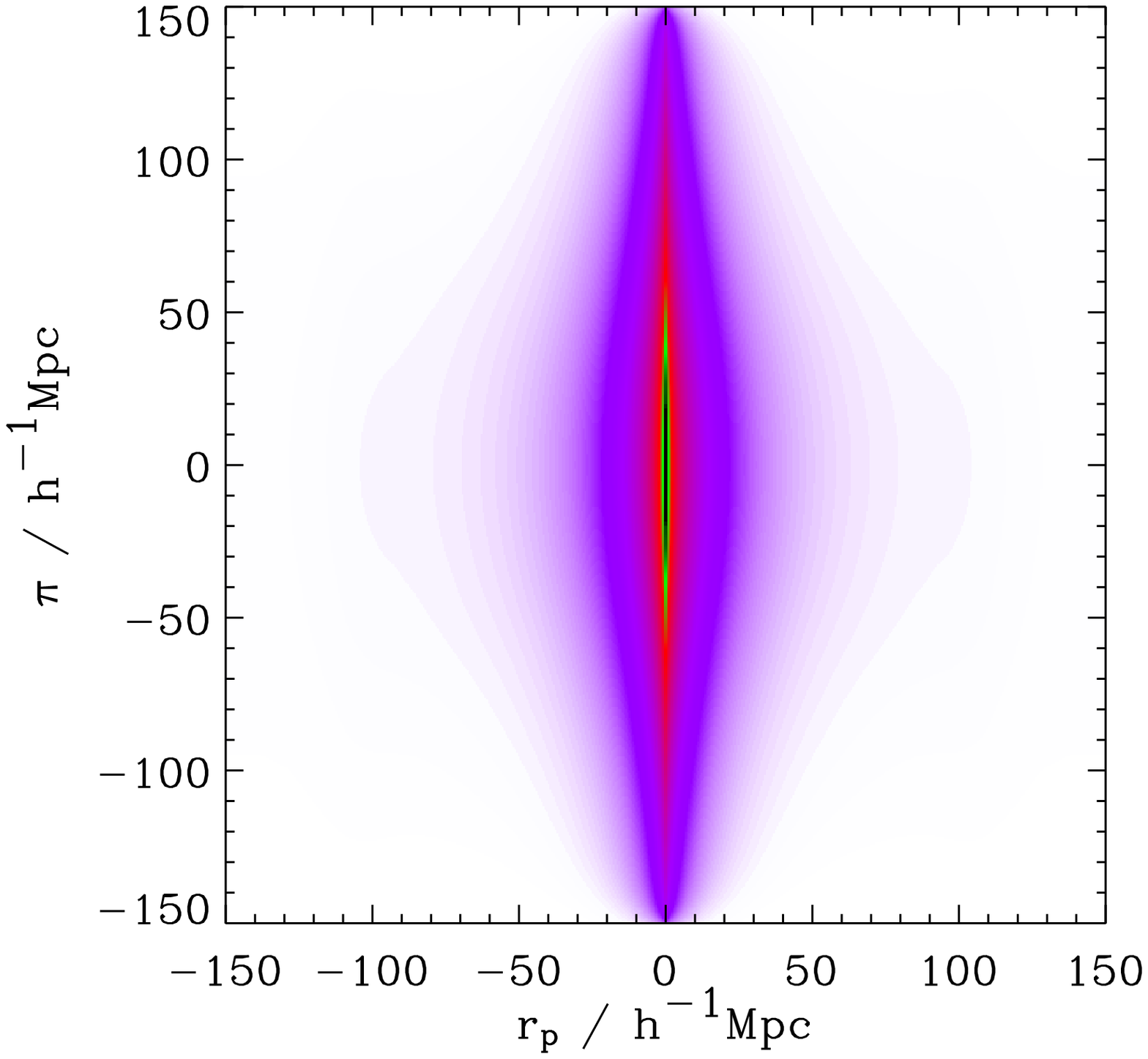,angle=0,clip=t,width=8.cm}}
\caption[ ]{The redshift-space correlation function $\xi(r_p,\pi)$ calculated using the halo model,
plotted as a function of transverse ($r_p$) and radial ($\pi$) pair separation. The data
from the first quadrant are repeated with reflection in both axes,
in order to clarify deviations from isotropy.  Upper panel: before convolution with
the pairwise redshift error distribution inferred from the sample of
COMBO-17 galaxies under consideration in this analysis, lower panel: after convolution.
\label{model}}
\end{figure}

\subsection{The effect of redshift errors on $\xi(r_p,\pi)$}\label{errorsonxi}

We now illustrate the effects of redshift errors, starting from
a model for the true $\xi(r_p,\pi)$. This is calculated using
the halo model, as described above in section~\ref{HaloModel},
together with a prescription for redshift-space distortions.
\citet{Seljak01} showed how to include redshift-space distortions
in the halo model, but it is also common to use the following simple
model for the ratio of redshift-space power to real-space power:
\begin{equation}
P_s/P_r = {(1+\beta\mu^2)^2 \over (1+k^2\mu^2\sigma_p^2/2)},
\end{equation}
where $\sigma_p$ is an effective pairwise velocity dispersion, $\mu$
is cosine of the angle between $\bf k$ and the line of sight, and
$\beta\equiv \Omega_m^{0.6}/b$ (e.g. \citealp{Ballinger96}).
This approach has the advantage that the redshift-space correlation
function can then be found analytically, apart from a radial convolution
(\citealp{Hamilton92}).
We used this prescription, taking $\sigma_p$ to be $\sqrt{2}$ times
the one-dimensional velocity dispersion calculated from the halo model.
Fig. \ref{model} shows a model $\xi(r_p,\pi)$, which contains
two well-known expected anisotropies: the isocorrelation contours
of $\xi(r_p,\pi)$ are stretched along the $\pi$ direction at small
separations, because of the effect of virialized velocity dispersions, and
compressed at large scales as a consequence of large-scale coherent
motions. The former effect is not clearly visible owing to the large
scale of the plot.

The effect of redshift errors on this redshift-space correlation
function is straightforward: it is a convolution in the radial direction.
This reflects the fact that $1+\xi$ is a ratio of the observed
and expected numbers of pairs of galaxies. The correlation function $\xi(r_p,\pi)$
becomes distributed more broadly along the $\pi$ axis, but the total
correlation signal is conserved -- this is why $w(r_p)$ is independent of redshift errors.
In order to model this process, we need a model for the convolving
function. This is simply deduced, because the classification scheme
automatically returns an estimate of the rms redshift error
for each galaxy (see \citealp{COMBOMain04}). Thus, given a pair
of galaxies $i$, $j$, with redshift errors $\sigma_{z_i}$ and
$\sigma_{z_j}$  the rms pairwise error is
$\sigma_{{\rm pair}_{i,j}}=\smash{(\sigma_{z_i}^2+\sigma_{z_j}^2)^{1/2}}$.
The signal from this pair is smeared by a Gaussian with this
width, so the overall convolving function is a sum of the Gaussians
corresponding to all pairs:
\begin{eqnarray}\label{pairwiseerrors}
f(\delta z)=\frac{1}{N}\sum_{\rm pairs}
(2\pi\sigma_{\rm pair})^{-1/2} \exp\left[-(\delta z/\sigma_{{\rm pair},n})^2/2\right]~,
\end{eqnarray}
where $N$ is the number of pairs $(i,j)$. After having transferred
the pairwise redshift error distribution into comoving distances we
can convolve $\xi(r_p,\pi)$ with Eq. (\ref{pairwiseerrors}).

The effect of this convolution is shown in the second panel in
Fig. \ref{model}.  The redshift-space correlations are now
heavily elongated in the radial direction, and some care
is needed in extracting the projected correlation signal.

\begin{figure}[h]
\centerline{\psfig{figure=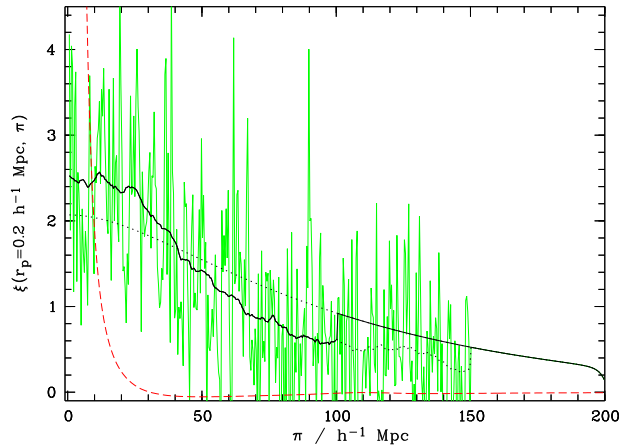,angle=270,clip=t,width=9.cm}}
\caption[ ]{An example of the model-fitting procedure used to determine
$w(r_p)$: The grey line shows the correlation function $\xi(r_p,\pi)$
of the red sequence galaxies as
a function of $\pi$ at  $r_p=0.2 \mpcoh$. The dashed line shows the
model correlation function in the absence of redshift errors.
The dotted and solid lines show the smoothed data  (the smoothing was
done by applying a box filter of length $40 \mpcoh$ along
the $\pi$-axis using the running average), and the model after the convolution with the
expected error distribution, 
with the solid part indicating where the integration is carried out.
\label{fitdata}}
\end{figure}

\subsection{Estimation of projected correlations}

The simplest strategy for carrying out the projection needed
in order to deduce $w(r_p)$ would be to integrate $\xi(r_p,\pi)$
over a very large radial range. Fig. \ref{model} suggests that
a maximum $\pi$ value of $150$ to $200 \mpcoh$ would be required
to capture all the signal. The problem with this strategy is that
the random noise in $\xi(r_p,\pi)$ is independent of $\pi$ at
a given $r_p$ (because the expected pair counts have a cylindrical
dependence $\propto r_p\, dr_p\, d\pi$). Thus, integration to
$\pi=200\mpcoh$ would yield a random error in $w(r_p)$ that is
$\sqrt{2}$ times larger than integration to $\pi=100\mpcoh$ -- but
the lower limit systematically misses part of the signal.

We have developed a strategy for solving this
problem, which depends weakly on some prior knowledge
of the likely form of the true clustering signal (after error convolution).
A model for $\xi(r_p,\pi)$ defines how the real-space signal $w(r_p)$
is spread out in $\pi$; we are only concerned with the shape of this
function, which is dominated by the convolution with the redshift
error distribution. Given this probability distribution in 
the $\pi$ direction, the amplitude of $w(r_p)$ can be estimated by
fitting a scaled version of our model $\xi(r_p,\pi)$ to the
data at the $r_p$ value of interest.

In practice, our model $\xi(r_p,\pi)$ will not be exact,  and we considered
the following compromise procedure for estimating $w(r_p)$ so that the result is robust.
For each $r_p$ value, we fit the amplitude of the (convolved) model
$\xi(r_p={\rm const},\pi)$ to the data. We then integrate
the data for $\xi(r_p,\pi)$ out to
$\pi=100\mpcoh$, from which point on we integrate the convolved model
out to infinity (see Fig. \ref{fitdata}).
This combines the exact measurement of $w(r_p)$ within $\pi_{\rm max}=100\mpcoh$ with
an estimate of the missing signal at larger $\pi$. Since this correction is typically
20\% of the overall signal, we do not need to estimate it very accurately.
In practice, the results from the 2-stage procedure were very similar to
the direct fitting method. This process is performed separately for
the red and blue galaxy samples, using the appropriate pairwise error distributions
and the final best-fitting halo-model $\xi(r_p,\pi)$ for the
unconvolved prediction. 

The width of the convolved model in Fig. \ref{fitdata} suggests that
the redshift errors yielded by the 
object classification scheme, which we used for the calculation of the
pairwise error distribution, may be slightly overestimated. In order to
estimate the effect on the projected correlation function, we
tried repeating the analysis with redshift errors scaled to $80$\% of the values
given in the object catalogues. This scaling gives the best fit to the data;
however, the resulting changes to $w(r_p)$ were small
compared to the random errors.

\subsection{Integral constraint}

The mean galaxy density is determined from the observed galaxy counts in each
field, which does not necessarily represent the the true density \citep{GrothPeebles77}.
The estimator will be on average biased low with respect to the
true correlation  by a constant ${\cal I}$:
\begin{eqnarray}
w_m(r_p)=w_t(r_p)-{\cal I}~,
\end{eqnarray}
where $w_t(r_p)$ is the true projected correlation function, $w_m(r_p)$ the
measurement. The integral constraint ${\cal I}$ is given by
\begin{eqnarray}
{\cal I}\simeq\frac{1}{S^2}\int{w_{\rm t}(r_p)\mathrm{d}^2
S_1\mathrm{d}^2 S_2}~,
\end{eqnarray}
where  $S$
is the physical area corresponding to the solid angle of the field
at the redshift under consideration. For the calculation of the
integral constraint, we assume that the 
three dimensional correlation function
$\xi(r)$ is to first approximation a power law: 

\begin{eqnarray}\label{powerlaw}
\xi(r)=\left(\frac{r}{r_0}\right)^{-\gamma}~.
\end{eqnarray}
Then the evaluation of equation \ref{projection} yields
\begin{eqnarray}\label{wrp}
w(r_p)=C r_0^\gamma r_p^{1-\gamma}~,
\end{eqnarray}
where $C$ is a numerical factor, which depends only on the slope $\gamma$:
\begin{eqnarray}\label{Gamma}
C=\sqrt{\pi}\;\frac{\Gamma((\gamma-1)/2)}{\Gamma(\gamma/2)}~.
\end{eqnarray}
If the {\it true} correlation function
is given by equation (\ref{wrp}), the measurement yields
\begin{eqnarray}
w_m(r_p)&=&C r_0^\gamma  r_p^{1-\gamma}-{\cal I}\label{integconstfit1}\\
&=&C r_0^\gamma  \left[r_p^{1-\gamma}-{\cal I}/(C r_0^\gamma )\right]~.\label{integconstfit2}
\end{eqnarray}
The true amplitude $C r_0^\gamma$ is not known, but
${\cal I}/(C r_0^\gamma )$ can be estimated by performing a Monte Carlo
integration (where we use the mean of the pair counts $\langle
RR\rangle$ at a projected distance $r_p$ of the
four fields):
 \begin{eqnarray}
\frac{{\cal I}}{C r_0^\gamma }=\frac{\sum{\left[\langle RR\rangle\cdot r_p^{1-\gamma}\right]}}{\sum{\langle RR\rangle}}\label{integconstfit3}~.
\end{eqnarray}
The true value of $C r_0^\gamma$ can be estimated by fitting equation
(\ref{integconstfit2}) to the data, taking the value of ${\cal I}/(C
r_0^\gamma)$ from equation (\ref{integconstfit3}). This value,
multiplied by the fitted amplitude $C r_0^\gamma$, yields the integral
constraint ${\cal I}$. The measurement can then be corrected for the integral constraint by
adding ${\cal I}$ to $w_m(r_p)$. 

This method yields estimates of
${\cal I}=0.14 \mpcoh$ for the red galaxies and ${\cal I}=0.33\mpcoh$ for the blue galaxies.
These values are negligible in comparison with the observed data for $r_p \la 20\mpcoh$, demonstrating that
the fields are large enough to deliver a fair sample.
As a cross-check, note that we expect ${\cal I}=200 \sigma^2$ (since we integrated over
$\Delta \pi=200\mpcoh$), where $\sigma^2$ is the fractional
variance in galaxy numbers between different realizations of our survey. With three fields,
$\sigma$ should be $\sqrt{3}$ times smaller than the field-to-field rms variation, so
our figures for ${\cal I}$ suggest 4.5\% and 7\% expected scatter in the numbers of
galaxies per field for red and blue galaxies respectively. This agrees well with the
numbers in Table~1.

\begin{figure}[h]
\centerline{\psfig{figure=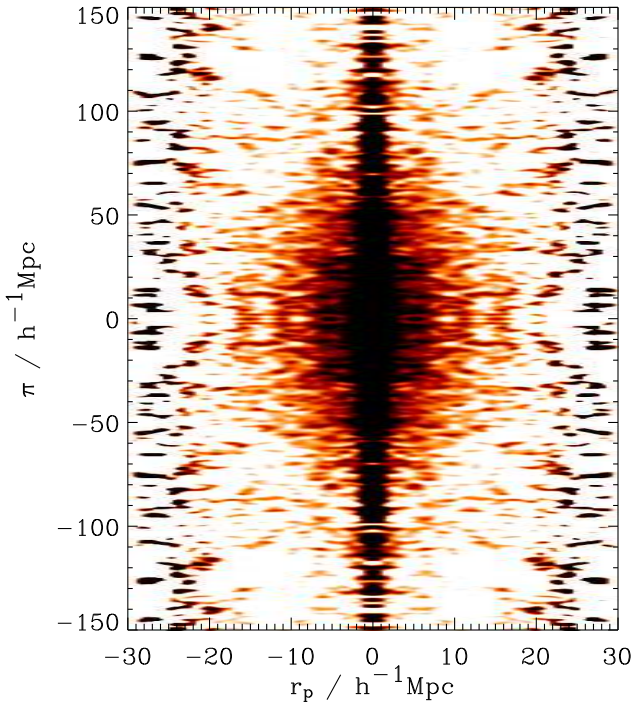,angle=0,clip=t,width=9.cm}}
\caption[ ]{$\xi(r_p,\pi)$ of all COMBO-17 galaxies with $0.4 < z < 0.8$, 
$I_{815} < 23$ and $M_B < -18$. Again the data from the first quadrant are
repeated with reflection in both axes. In the transverse direction the
pair separations accessible for the analysis are limited by the COMBO-17 field of view.
\label{COMBOallGals}}
\end{figure}

\subsection{Error analysis}

Finally, there is the crucial issue of setting realistic error bars on
our correlation estimates. The three COMBO-17 fields measure 
$\sim 31'\times 30'$ each and are thus large enough to carry out a
jack-knife analysis. We divide each field into four quadrants, and then
calculate the correlation function $w(r_p)$ (including the integral constraint)
for twelve realisations of the data, each time
omitting one of the quadrants. The variance in $w$ is then given approximately by
\begin{eqnarray}
\sigma^2=\frac{N-1}{N}\sum_{i=1,N}\left[\langle w(r_p)\rangle-w_i(r_p)\right]^2~,
\end{eqnarray}
where $N=12$ is the number of realisations of the data (e.g. \citealp{Scranton02}).

In order to check for cross-correlations between the data points, we can
extend the jack-knife method in the obvious way to estimate
the covariance between different bins, $\sigma^2_{ij}$. The natural
way to express this is as a correlation coefficient matrix:
$r_{ij}\equiv\sigma^2_{ij}/\sigma_{i}\sigma_{j}$.
Results in this form are presented below.
 
\begin{figure}[h]
\centerline{\psfig{figure=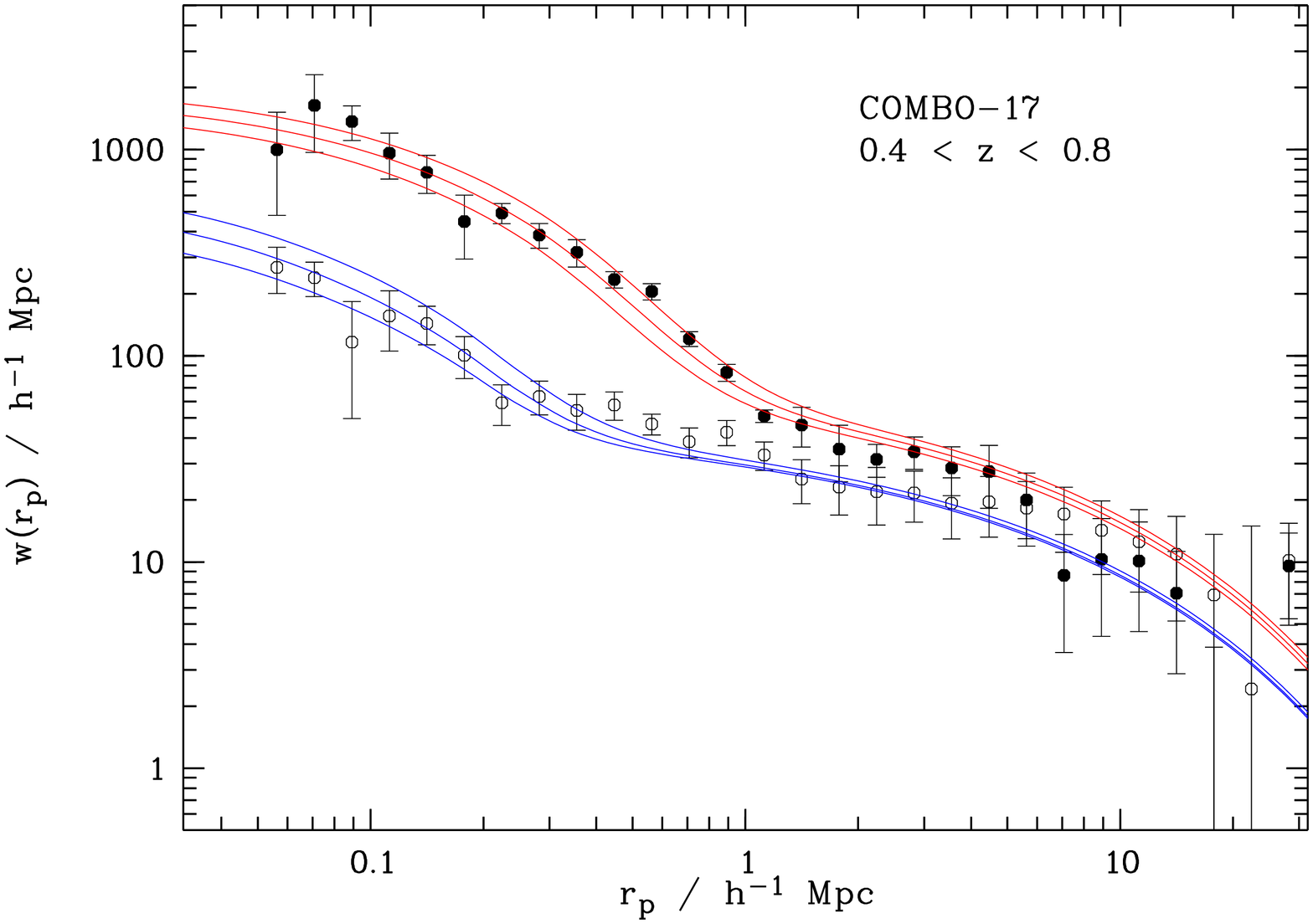,angle=270,clip=t,width=9.cm}}
\centerline{\psfig{figure=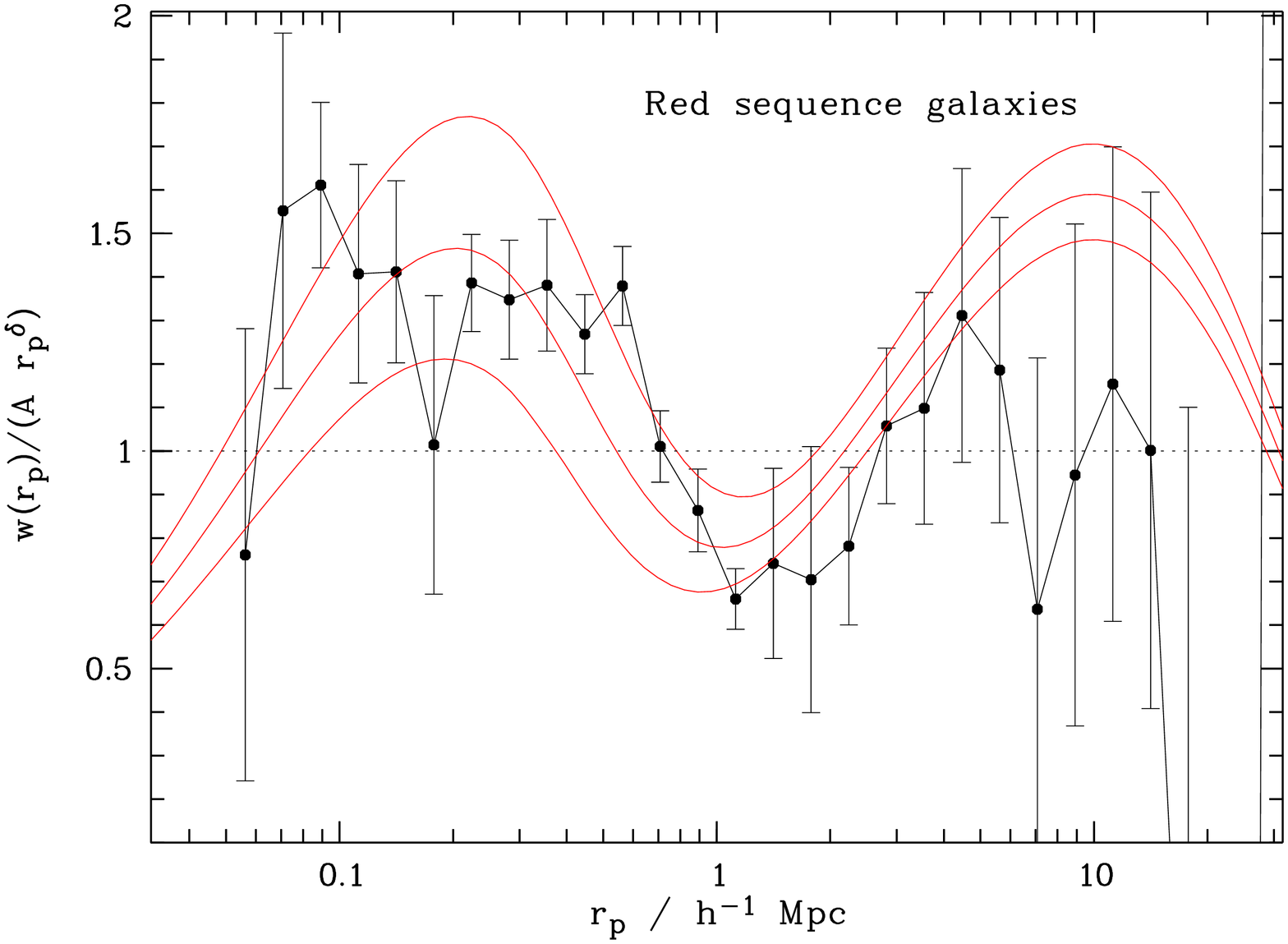,angle=270,clip=t,width=9.cm}}
\centerline{\psfig{figure=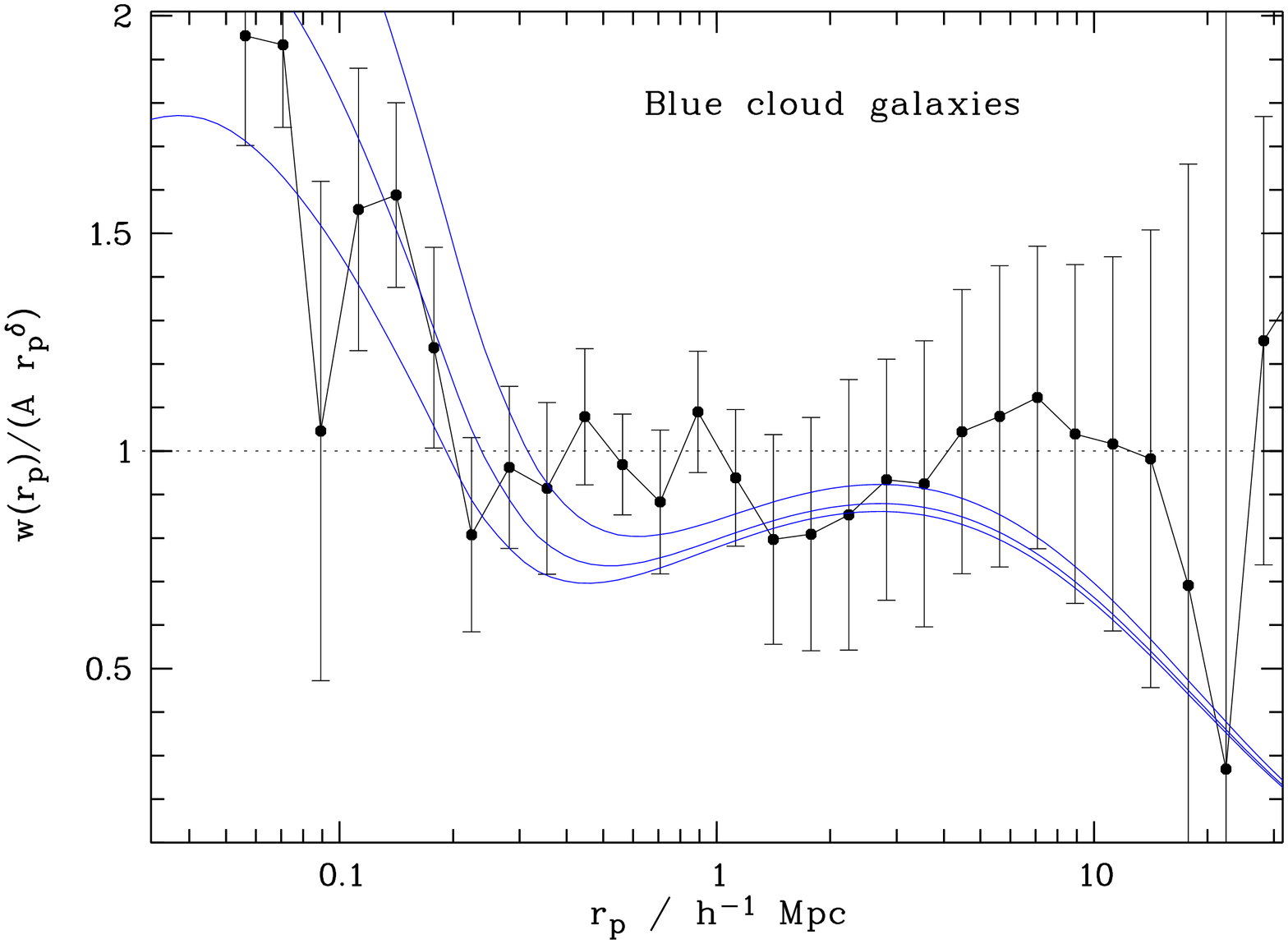,angle=270,clip=t,width=9.cm}}
\caption[ ]{The projected correlation function $w(r_p)$ for 
the COMBO-17 data, divided into red and blue galaxies. 
The lines show predictions from the halo model with $\alpha=0.45, 0.5, 0.55$
for red sequence and $\alpha=0.15, 0.2, 0.25$ for blue cloud galaxies, respectively,
where increasing $\alpha$ corresponds to increasing clustering.
The lower panels show an expanded view of the data,
divided by the best-fitting power
law (fitted in the range $\logten r_p < 1.1$).
\label{wrpredandbluec17}}
\end{figure}

\section{The clustering of the COMBO-17 galaxies}\label{Results}

\subsection{Results}

We calculated $\xi(r_p,\pi)$ for all COMBO-17 galaxies in the
redshift range $0.4 < z < 0.8$ with $I$-band magnitudes $I < 23$ 
and absolute restframe $B$ band luminosities $M_B < -18$.
We used the estimator invented by \citet{LandySzalay93}. 
An angular mask for the survey was derived by censoring the
surroundings of bright stars in the fields. The same mask was applied to a random
catalogue consisting of $30\,000$ randomly distributed galaxies, each of which
was assigned a redshift taken randomly from the real data,
where the three fields were put together in order to smooth the
redshift distribution. Using a smoothed form of the empirical redshift
distribution did not yield a significant change in the results.

The resulting $\xi(r_p,\pi)$ is shown in Fig. \ref{COMBOallGals}.
The field of view of the COMBO-17 fields limits the pair
separations accessible for the analysis, so in the transverse direction
there is of course no signal at separations larger than the physical
distance corresponding to the diagonal diameter of the fields.

For each object, we have an estimate of the redshift and the restframe colours
and luminosities; it is therefore possible to divide the sample into two distinct 
colour classes as described earlier.
For both samples we calculated $w(r_p)$ as described in section
\ref{Method}, correcting for the integral
constraint ${\cal I}$, and the influence of the redshift errors.
These results are shown in Fig.~\ref{wrpredandbluec17}.

\subsection{Fitting the halo model}

Fig.~\ref{wrpredandbluec17} also shows predictions from the halo model,
varying the single occupation-number parameter $\alpha$, and
choosing the cutoff $M_c$ so as to match the observed comoving
densities of $0.004h^3{\rm Mpc}^{-3}$ (red) and
$0.012h^3{\rm Mpc}^{-3}$ (blue). It
is apparent that there is greater sensitivity to $\alpha$
at small separations, and that once $\alpha$ is fixed from
the data there, there is little freedom at large separations, where
the data and the model match satisfyingly well. The preferred
values are approximately $\alpha=0.5$ for the red population and
$\alpha=0.2$ for blue galaxies. These figures correspond
to cutoff masses of respectively
$M_c=10^{12.15} h^{-1} M_\odot$ and $M_c=10^{11.50} h^{-1} M_\odot$.
As discussed earlier, a more meaningful way of casting these numbers
may be to apply the HOD model to the halo mass function, to calculate
the effective halo mass, weighting by galaxy number. These figures
come out as $M_{\rm eff}=10^{13.21} h^{-1} M_\odot$ and $M_{\rm eff}=10^{12.52} h^{-1} M_\odot$
respectively.

Fig. \ref{wrpredandbluec17} also shows a magnified view,
with the measured correlation functions and the
corresponding best-fitting models both divided by a power-law fit (fitted
in the range $\logten r_p < 1.1$), the slope and amplitudes of which
are given in Table \ref{powerlawtab}. The data points do not scatter
arbitrarily around the power-law fit, but show systematic deviations.
For the red galaxies, there is a marked dip around
$r_p\simeq 1.5 \mpcoh$; the blue galaxies are closer to a power
law, but with a relatively abrupt step at $r_p \simeq 0.2 \mpcoh$.
Both these features are impressively well accounted for by the
halo model predictions, especially when it is considered that there
is only one free parameter.

\begin{table}[h]
\centering
\caption[ ]{The amplitudes, slopes and resulting correlation lengths $r_0$
of the best fitting power law
($w(r_p)=A r_p^\delta$, fitted  for $\logten r_p< 1.1$), for 
the total sample, and divided into red
sequence and blue cloud galaxies.\\\label{powerlawtab}}
\begin{tabular}{c c c }
$A_{\rm tot}$&$\delta_{\rm tot}$&$r_0^{\rm tot}$\\ \hline
$53.98\pm^{3.86}_{3.50}$&$-0.57\pm0.02$&$4.71\pm^{0.26^{\strut}}_{0.26_{\strut}}$\\ 
$A_{\rm red}$&$\delta_{\rm red}$&$r_0^{\rm red}$\\ \hline
$86.28\pm^{4.86}_{4.50}$&$-0.94\pm0.03$&$5.39\pm^{0.30^{\strut}}_{0.28_{\strut}}$\\ 
$A_{\rm blue}$&$\delta_{\rm blue}$&$r_0^{\rm blue}$\\ \hline
$37.09\pm^{2.58}_{2.41}$&$-0.45\pm0.03$&$3.64\pm^{0.25^{\strut}}_{0.24_{\strut}}$\\
\end{tabular}
\end{table}

It is interesting to compare our results with those of the VVDS project
(\citealp{VVDS}). They give results to a similar depth for two fields, although not divided by
colour, with a total of 7155 redshifts over 0.61~deg$^2$. Their redshift bins are not
identical, but they quote $r_0=2.69^{+0.53}_{-0.59}\mpcoh$ and
$\gamma=1.71^{+0.18}_{-0.11}$ at $\langle z \rangle=0.6$
and $r_0=4.55^{+1.25}_{-1.56}\mpcoh$ and
$\gamma=1.48^{+0.28}_{-0.15}$ at $\langle z \rangle=0.7$.
The latter figure is from the CDFS, which is one of our fields,
and we have checked that our figure for this field alone agrees well
with the VVDS, as it does for our other fields. The VVDS $2^h$ field
thus gives a somewhat lower clustering strength; this may be
because the VVDS sample in that field is about 0.5 mag. deeper than the one
studied here, plus the fact that the VVDS analysis has no lower limit
in luminosity. The ratio in $r_0$ between the VVDS $2^h$ field and
our overall result is $1.75 \pm 0.21$, so a factor 1.3 from
luminosity-dependent clustering would be 
required in order to make the results statistically consistent.

\begin{figure}[ht]
\centerline{\psfig{figure=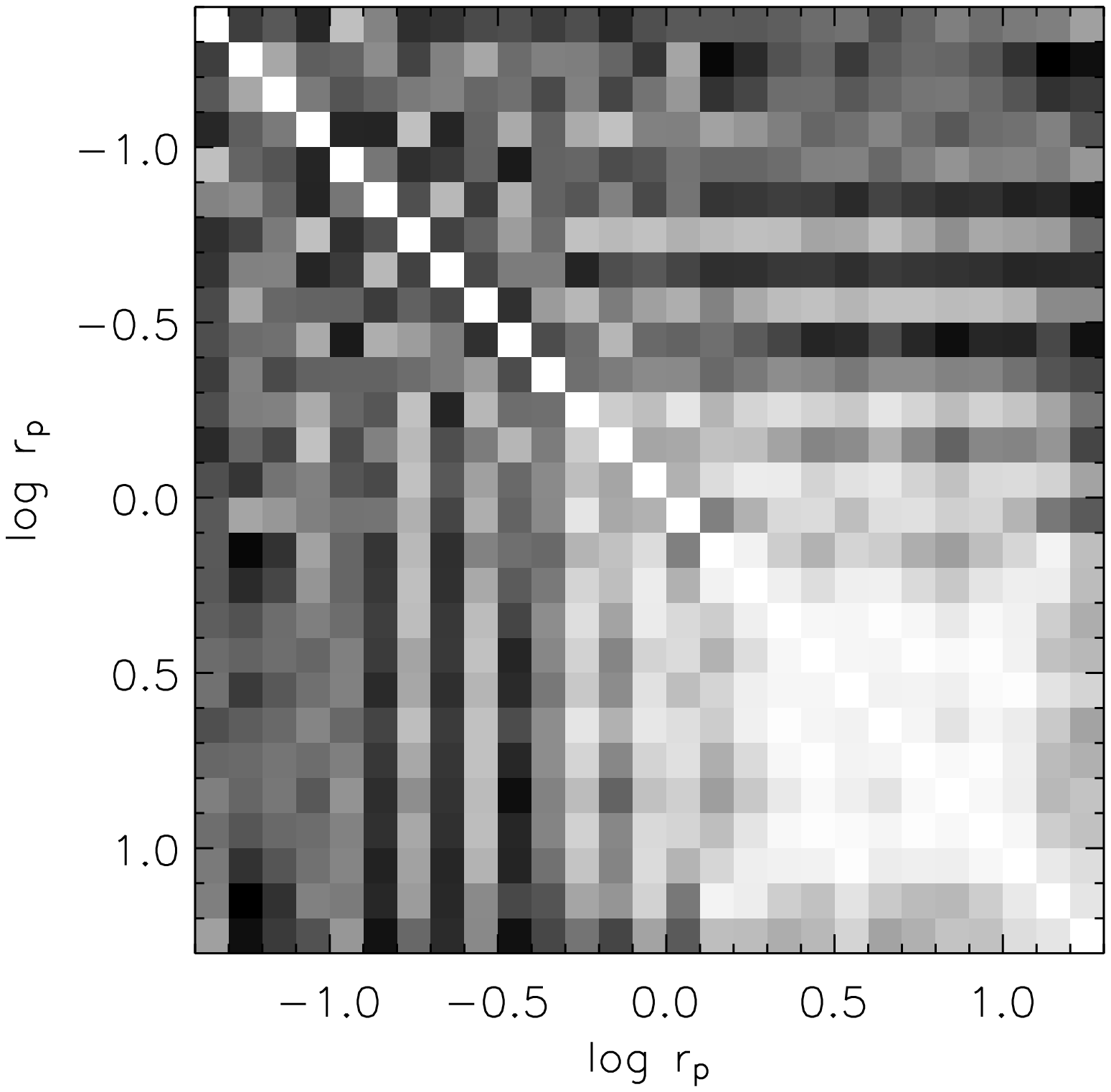,clip=t,width=8.cm}}
\centerline{\psfig{figure=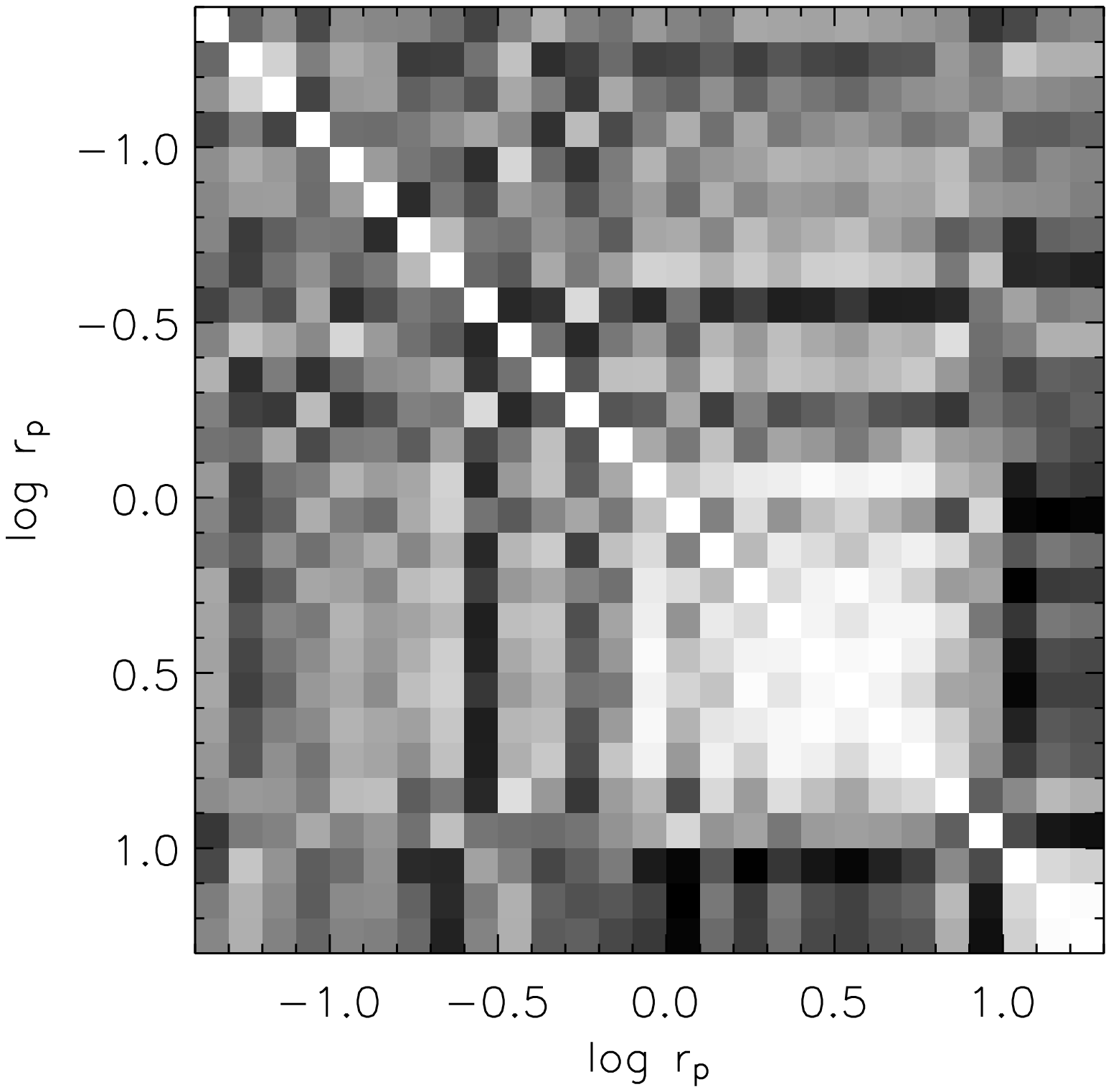,clip=t,width=8.cm}}
\caption[ ]{The correlation matrices for red (upper panel) and blue
(lower panel)
galaxies, for $-1.4 < \logten r_p < 1.3$. The apparent strong correlation between the large scale data
points (lower right corner) is due
to the integral constraint.
\label{correlationmatrix}}
\end{figure}

\begin{figure}
\centerline{\psfig{figure=3626f12.ps,clip=t,width=7.cm}}
\centerline{\psfig{figure=3626f13.ps,clip=t,width=7.cm}}
\caption[ ]{The likelihood contours for the two free
parameters $\alpha$ and $\sigma_8$. Upper panel: red galaxies;
lower panel: blue galaxies. 
Contours are shown at the 
Gaussian equivalent of the following confidence levels: 68\% 1-parameter,
68\%, 95\%, 99\%, 99.9\%, 99.99\% 2-parameter, i.e. $\Delta \ln {\cal L}= 0.5$,
1.1, 3.0, 4.6, 6.9, 9.2.
\label{chiq_c17}}
\end{figure}

\begin{figure}[ht]
\centerline{\psfig{figure=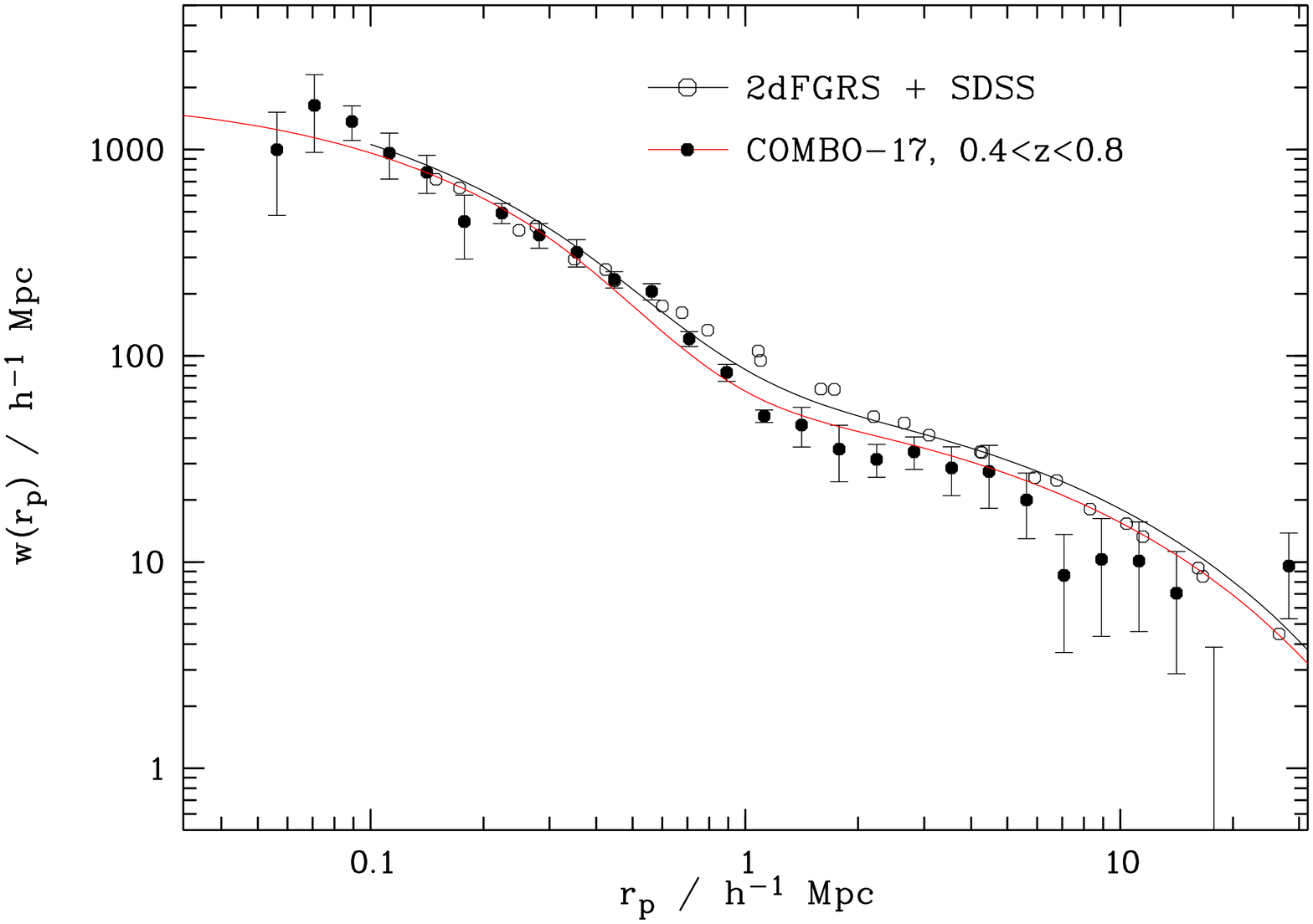,angle=270,clip=t,width=9.cm}}
\centerline{\psfig{figure=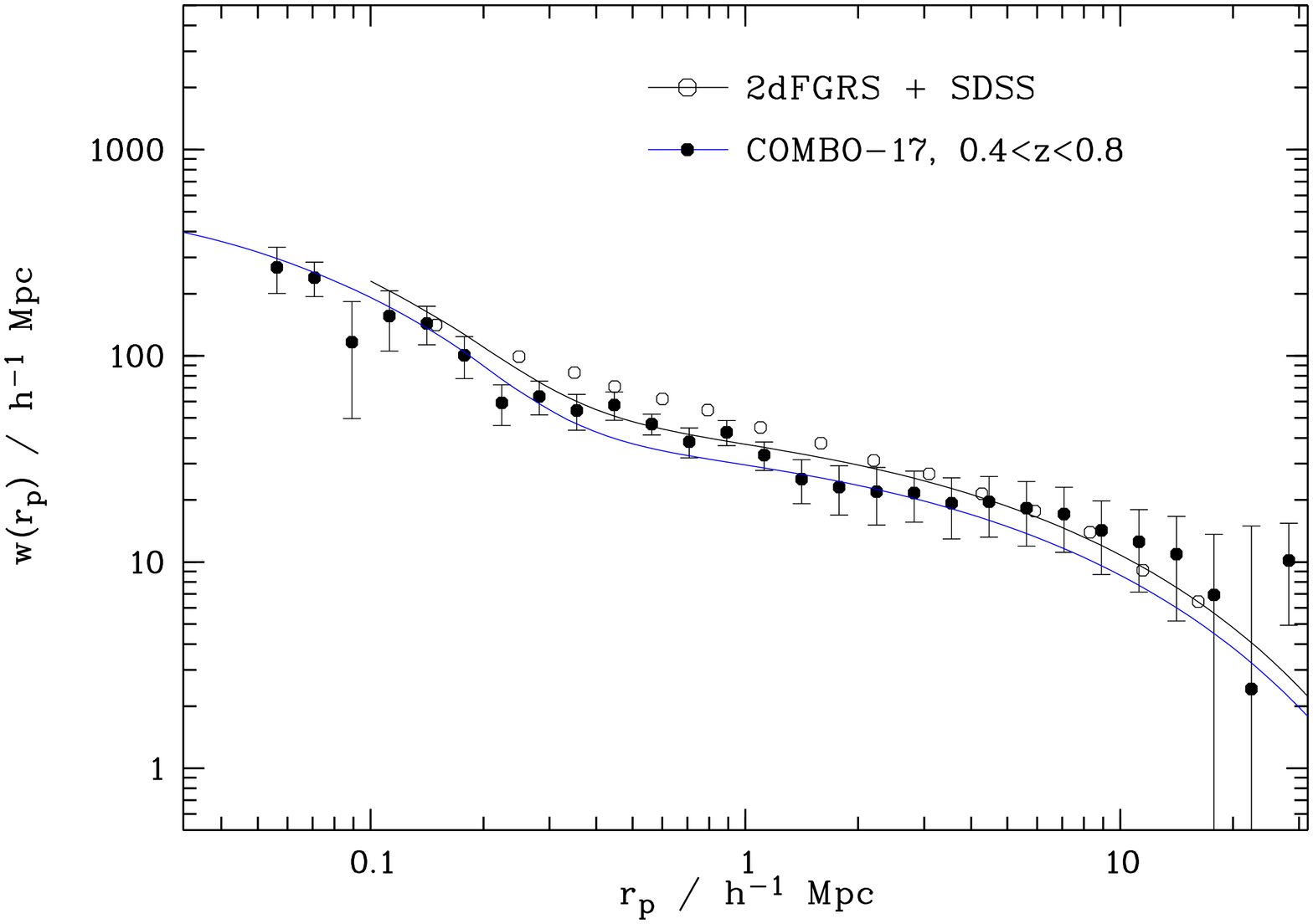,angle=270,clip=t,width=9.cm}}
\caption[ ]{The projected correlation function in the
redshift bin $0.4< z < 0.8$ (filled circles) in comparison with the local data
(open circles). The upper panel shows red sequence
galaxies, the lower panel is for blue cloud galaxies. The lines show the prediction
of the halo model, which are higher in the $z=0$ case.
\label{z0_z6_red}}
\end{figure}

\begin{figure}[ht]
\centerline{\psfig{figure=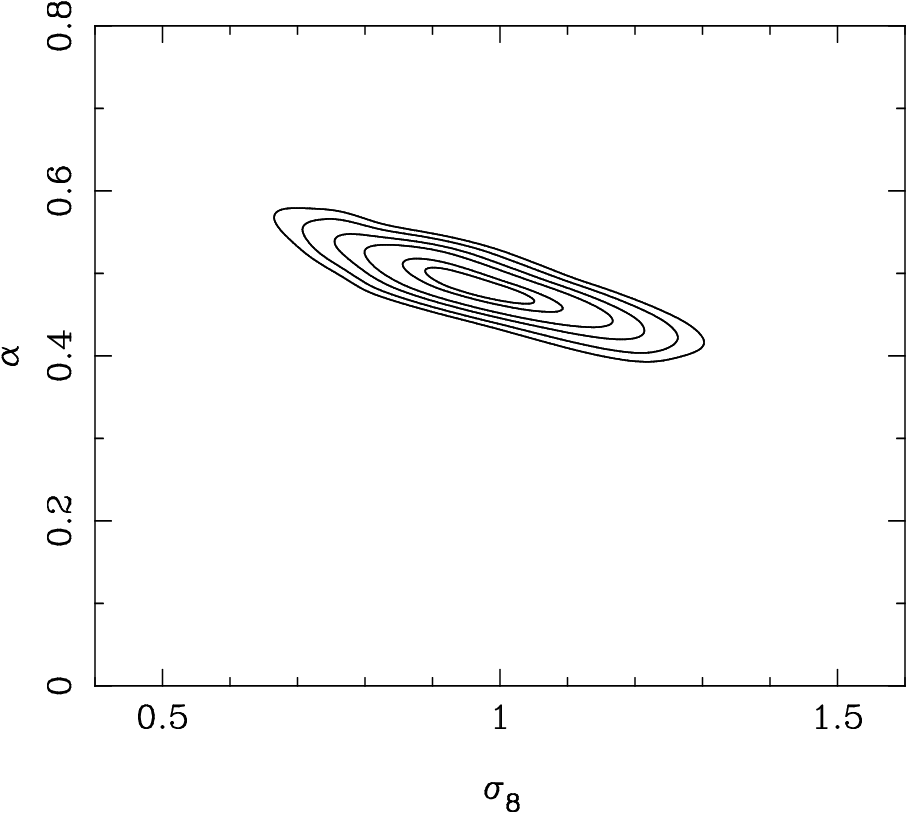,clip=t,width=7.cm}}
\centerline{\psfig{figure=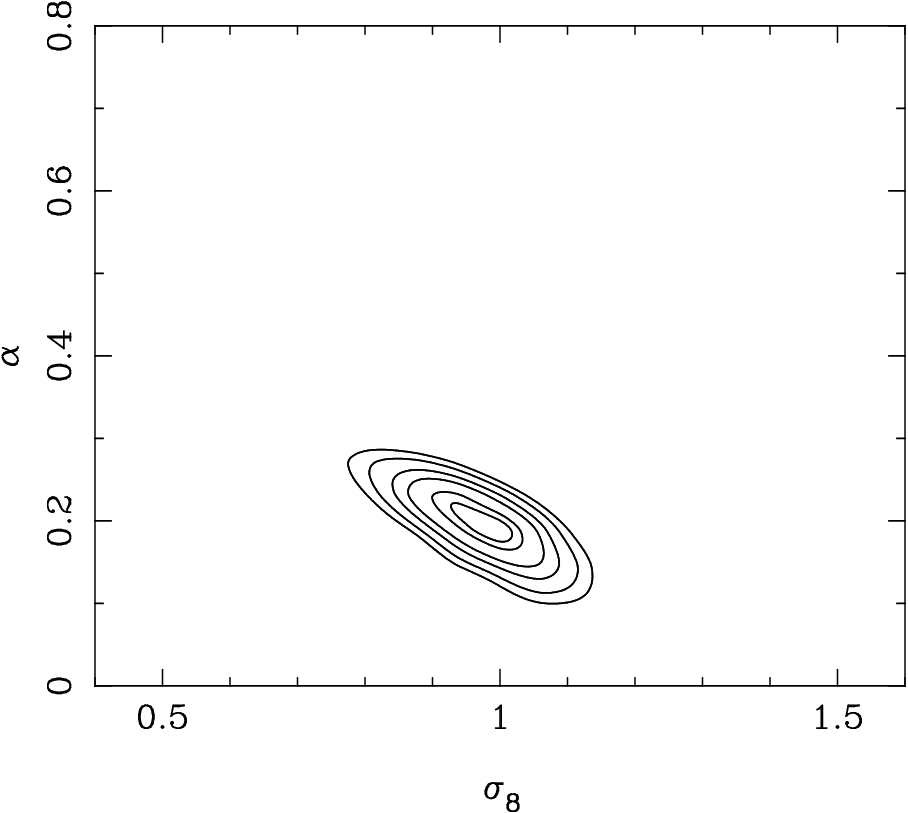,clip=t,width=7.cm}}
\caption[ ]{The likelihood contours for the two free
parameters $\alpha$ and $\sigma_8$, now fitting the $z=0$ data.
Upper panel: red galaxies; lower panel: blue galaxies.
Contours are shown at the 
Gaussian equivalent of the following confidence levels: 68\% 1-parameter,
68\%, 95\%, 99\%, 99.9\%, 99.99\% 2-parameter, i.e. $\Delta \ln {\cal L}= 0.5$,
1.1, 3.0, 4.6, 6.9, 9.2.
\label{chiq_z0}}
\end{figure}

We now want to quantify the agreement between model and data in more detail,
using the jack-knife error estimates. This would be straightforward if the
covariance matrix were diagonal, and if the model was exact. In
practice, there is some degree of correlation in the data, and the
simple halo model used here may be expected to have some systematic
deviations with respect to ideal data. 
The correlation matrices in Fig. \ref{correlationmatrix} indeed show a
strong correlation between the large-scale data
points (large matrix indices, lower right corner). This is due
to the integral constraint, which is included in the calculation of
the jack-knife realisations: all large-scale data points
are offset by the same amount and thus become correlated.
This issue is not too serious, since the errors in this regime
are in any case large. We therefore ignore the large-scale
points at $r_p>10\mpcoh$ and treat the remaining data as independent.
Even so, our approximate model is not guaranteed to deliver a
perfect fit. In order to achieve a formally acceptable value of
$\chi^2$ for the fit, we added in quadrature an error of
5\% to the errors on $w$ for red galaxies. When fitting the $z=0$ data, as discussed
below, a covariance matrix was not available, and the formal errors
are in any case small. In this case, we therefore took what is effectively
a least-square approach, which required an effective error of 10\% in
both blue and red galaxies.

In performing the fitting, it is interesting to consider variations in
both the power-law index $\alpha$ of
the HOD (equation \ref{HODeqn}), and in the normalization $\sigma_8$.
We emphasise that $\sigma_8$ is the zero-redshift value, which is connected
to the degree of inhomogeneity at $z=0.6$ by the growth factor
predicted by the cosmological model (a change in linear density contrast
by a factor 1.32).
Fig. \ref{chiq_c17} shows the likelihood contours for these two free
parameters $\alpha$ and $\sigma_8$.
The preferred values and marginalized rms errors are
$\alpha=0.56\pm0.03$ and $\sigma_8=0.84\pm0.08$ for
the red sequence galaxies and  $\alpha=0.16\pm0.03$ and
$\sigma_8=1.19\pm0.09$ for the blue population.

These independent estimates of
$\sigma_8$ from the red and blue populations are both close to
our default value of 0.9. The agreement is not perfect, and the
difference in $\sigma_8$ is formally $3\sigma$, but the mean value of
$\sigma_8=1.02 \pm 0.17$ is certainly plausible. 
Given the simplicity of the HOD model, this is a satisfying result.

\subsection{Robustness of the results}

In view of the tension between the normalization inferred from the
red and blue galaxies, and in the light of the preference for a
low normalization of $\sigma_8\simeq0.75$ from the 3-year WMAP data
\citep{Spergel06}, it is important to discuss the extent to
which our result is rendered uncertain by simplifying assumptions
in the modelling.

The halo model is an idealized approximation in many ways,
and one issue in particular has generated considerable discussion recently.
It has been shown that the clustering of dark matter haloes
is dependent on the halo formation time
\citep{ShethTormen04a,ShethTormen04b,Gao05,Wechsler05,Harker06,Reed06},
and the question is what impact this has on halo model
calculations. To some extent, the effect is already included in
the halo-model formalism: when the cosmic density field
is smoothed on a given mass scale, the clustering of peaks in the
smoothed field is well known to increase with peak height, i.e.
with formation redshift (\citealp{Kaiser84}). More massive systems are
more strongly clustered for the same reason: only the rarest peaks
exceed the threshold for collapse when the variance in filtered density is low.
When we use the standard expression for bias as a function of
mass, this averages over all systems that have collapsed by the
present: the dependence of clustering on 
formation time will thus have no effect on predicted galaxy
properties if the occupation numbers are purely a function of halo mass.
However, it seems reasonable that the occupation number for a given mass will
in fact depend to some extent on collapse redshift (e.g. more red
galaxies in a halo that collapses early). At a minimum,
the age-clustering effect will then contribute to a stochastic aspect
of the occupation number, so that there is some scatter in $N$ at
a given $M$. More seriously, it can also bias the mean clustering
compared to all haloes of that mass. The influence of these effects
on halo-model predictions remains to be explored, and this task is
beyond the scope of the present paper. Some work along these
lines has been done by \citet{Zentner05} and by \citet{Croton06},
where the halo contents in a simulation are scrambled between
all haloes of the same mass, thus destroying any correlations with
collapse redshift. \citet{Zentner05} did this for subhaloes, but
\citet{Croton06} considered the case of most direct interest, which
is semianalytic galaxy populations. They do detect systematic
shifts in correlation amplitude, but
for the luminosities of interest here these are no
larger than 10\%. Such shifts are not important in comparison with
the COMBO-17 measuring errors, but this is clearly an issue that
should be looked at in more detail, especially as the accuracy in
measuring high-redshift clustering improves.

\begin{table}
\centering
\caption[ ]{The inferred values
of $\sigma_8$ from COMBO-17 red and blue galaxies,
showing the effect of variations in some of the assumptions
in the halo-model calculation. We also give the corresponding values of
$\alpha$, the power-law slope in the occumaption model, plus formal
values of $\chi^2$ for the best-fitting models (on 20 data points).
\\\label{varymodtab}}
\begin{tabular}{c|cc|cc|cc}
Halo Model& $\sigma_8^{\rm red}$& $\sigma_8^{\rm blue}$& $\alpha_{\rm red}$& $\alpha_{\rm blue}$
& $\chi^2_{\rm red}$& $\chi^2_{\rm blue}$
\\ \hline
Standard & 0.84 & 1.19 & 0.56 & 0.16 & 26.2 & \phantom{0}9.7 \\
$\Omega_m=0.2$ & 0.80 & 1.07 & 0.52 & 0.16 & 38.0 & 16.1 \\
$h=0.6$ & 0.90 & 1.15 & 0.51 & 0.14 & 33.0 & 10.4 \\
$1+\delta_v=300$ & 0.90 & 1.02 & 0.55 & 0.18 & 43.1 & \phantom{0}9.6 \\
Zheng $N(M)$ & 0.85 & 0.98 & 0.96 & 0.56 & 46.4 & \phantom{0}9.4 \\
\end{tabular}
\end{table}

Other degrees of freedom in the halo model are more easily investigated, and 
we summarise some tests here, the results of which are presented in
Table~\ref{varymodtab}. We show the impact on the values of $\sigma_8$
inferred from red and blue galaxies by (a) varying cosmological parameters;
(b) varying parameters internal to the halo model; (c)
varying the occupation number prescription. In the first
category, we see that the Hubble parameter has very little effect,
but that $\sigma_8$ increases with $\Omega_m$ very roughly as
$\Omega_m^{0.3}$ to $\Omega_m^{0.5}$, with a larger sensitivity for the
blue galaxies. In the second category, we considered altering the
assumed density contrast for a virialized halo from the usual figure
of 200 to a slightly larger number, which is sometimes assumed
in a low-density model. Finally, we modify our simple power-law
$N(M)$ to something that resembles more closely the prescription
used by e.g. \citet{Zheng05}: $N=1$ between $M_c$ and $10M_c$,
rising as $M^\alpha$ for smaller $M$ (we considered $\alpha \le 1$).
We also include the formal $\chi^2$ values for some of these alternatives.
It appears that the standard model provides the best fit, especially
to the red galaxies. However, bearing in mind the simplified nature of the
halo model, these differences should not be given
high weight; it is more interesting to concentrate on the
robustness of the best-fitting parameters.

The overall conclusion of these tests is that plausible variations of
some of the degrees of freedom in the modelling can alter $\sigma_8$
by 10 to 20\%, and in a way that changes the consistency between
red and blue results by a similar amount. We therefore conclude
that the modelling is working as well as could have been 
expected, and that there is no need to be concerned by either
the internal red-blue tension or by the 1.6$\sigma$ discrepancy
with the WMAP $\sigma_8$. We now turn to the comparison with $z=0$;
some of the systematics in the analysis should be common to
all redshifts, so we should hope for a good level of consistency
between measurements based on data from different epochs.

\subsection{Comparison with local clustering}

As a local comparison, we considered $w(r_p)$ for a combined set of red and
blue galaxies taken from both Sloan Digital Sky Survey (SDSS;
\citealp{York00}) and 2dFGRS \citep{Colless01} data. The sample has
been divided into red and blue by either the bimodality of the rest-frame colour
distribution, or spectral type, in a way that should compare reasonably
well with the COMBO-17 classification. We use the flux-limited 2dFGRS
results of \citet{Hawk03}, and the $-19 > M_r >-20$ results of
\citet{Zehavi05}, which have closely comparable amplitudes.
Fig. \ref{z0_z6_red} shows a comparison of the local and the high redshift sample.
What is apparent here is that the main difference between
the COMBO-17 sample at $z=0.6$ and the local data is in the
{\it shape\/} of the correlation function, with almost identical
amplitudes at small scales, but a difference of nearly a factor two
for $r_p > 1 \mpcoh$. 

As shown in Fig. \ref{z0_z6_red}, the halo model is capable of
accounting for these differences.
Fig. \ref{chiq_z0} repeats the model-fitting exercise for the $z=0$ data, and
the preferred values and marginalized rms errors are
$\alpha=0.49\pm0.02$ and $\sigma_8=1.03\pm0.07$ for
the red sequence galaxies and $\alpha=0.23\pm0.02$ and
$\sigma_8=1.00\pm0.05$ for the blue population.
These values of $\alpha$ are slightly smaller than 
those obtained from COMBO-17; they correspond
to cutoff masses of respectively
$M_c=10^{12.20} h^{-1} M_\odot$ and $M_c=10^{11.50} h^{-1} M_\odot$,
or effective halo masses
$M_{\rm eff}=10^{13.50} h^{-1} M_\odot$ and $M_{\rm eff}=10^{12.80} h^{-1} M_\odot$
respectively.
Compared to the $z=0.6$ results, $M_c$ has not changed much, but
$M_{\rm eff}$ has increased by about a factor 2. This makes sense in terms
of hierarchical growth: the minimum dark-matter mass needed to assemble
a galaxy-sized amount of baryons should be invariant, but such
haloes inevitably merge into larger systems as time progresses.

The mean value of $\sigma_8=1.01\pm0.04$ from the $z=0$ data agrees very well with
$\sigma_8=1.02 \pm 0.17$ from COMBO-17. This agreement
assumes hierarchical growth in the halo mass function between
$z=0.6$ and the present, without which the inferred values of
$\sigma_8$ would have been expected to differ by a factor 1.3.

\section{Summary \& conclusions}\label{Discussion}

Using a sample of 10\, 360 galaxies with photometric redshifts
from the COMBO-17 survey, we have investigated in some detail the 
shape of the correlation function at redshift $\langle z \rangle  \simeq 0.6$.
We have shown for the first time that the two-point correlation
functions of both red sequence 
galaxies and blue cloud galaxies at this redshift display deviations
from a power law,  
analogous to the deviations seen at low redshift (\citealp{Hawk03,Zehavi04}).

We have compared these observations to the predictions of a simple halo model,
and find a good fit. It appears that the COMBO-17 data allow us to identify
the point of transition between 1-halo clustering and 2-halo clustering,
as was done at $z=0$ by \citet{Zehavi04}. The implication is that the
red and blue galaxies at $\langle z \rangle=0.6$ inhabit haloes of typical effective mass
$M_{\rm eff}=10^{13.2} h^{-1} M_\odot$ and $M_{\rm eff}=10^{12.5} h^{-1} M_\odot$
respectively.

We have also allowed the zero-redshift normalization of the power spectrum, $\sigma_8$,
to be a free parameter in this analysis.
Impressively, both red and blue subsets imply a consistent local normalization
of the power spectrum: $\sigma_8 \simeq 1$. This figure is close to the value inferred
by independent means using CMB and gravitational lensing (e.g. \citealp{Refregier03}),
and certainly within the tolerance expected from the inevitable systematics
associated with the simple modelling that we have used in this
first analysis of galaxy clustering at intermediate redshifts.
This consistency is obtained by assuming that the dark halo mass
function grows in a standard hierarchical fashion so that the normalization
at $z=0.6$ is approximately 30\% lower than today. Our results 
amount to a verification that growth of this order has occurred.

We intend to expand this
work to higher redshifts using COMBO-17{+4}, the NIR extension of
COMBO-17, for which observations are currently being
carried out using the $2k\times2k$ Omega2000 camera at the
3.5m-telescope on Calar Alto, Spain.
Combining the existing
optical data base from COMBO-17 with NIR observations in one broad and
three medium band filters (covering the wavelength range from 1040 to
1650\,nm), we expect to obtain
$\simeq 4200$ galaxy  redshifts with an accuracy of $\sigma_z/(1+z)=0.02$ up to $z=2$.
This much longer baseline in cosmic time will allow us to observe
much larger evolution of halo masses, testing the idea of hierarchical
growth back to a time close to the formation of luminous galaxies.

\begin{acknowledgements}
S. Phleps acknowledges financial support
by the SISCO Network provided through the European
Community's Human Potential Programme under contract
HPRN-CT-2002-00316.
JAP was supported by a PPARC Senior Research Fellowship.
CW was supported by a PPARC Advanced Fellowship.
We thank Luigi Guzzo for helpful comments.
We thank the anonymous
referee for helpful comments and suggestions.
\end{acknowledgements}

\bibliographystyle{aa}


\end{document}